\documentclass[MSNbibl,nameyear,dvips]{arxstspdf}
\usepackage{flushend}
\usepackage{stfloats}
\usepackage{graphicx}
\volume{28}
\issue{3}
\pubyear{2013}
\firstpage{447}
\lastpage{463}
\doi{10.1214/12-STS411} %kopijuoti is PTS
\makeatletter

\setattribute{abstract}{width}{410pt}

\newcommand{\noinbf}{\textbf}

\makeatother

\begin{document}
\begin{frontmatter}

\title{A Conversation with Stephen E. Fienberg}%\thanksref{T1}
% kai straipsnis turi susijusiu diskusiju ir rejoinder'iu
%rejoinder at \relateddoi{r}{10.1214/00-STSXXXX}.}
\runtitle{A Conversation with Stephen E. Fienberg}

\begin{aug}
\author[a]{\fnms{Miron L.} \snm{Straf}\corref{}\ead[label=e1]{mstraf@nas.edu}}
\and
\author[b]{\fnms{Judith M.} \snm{Tanur}\ead[label=e2]{Judith.Tanur@stonybrook.edu}}
\runauthor{M. L. Straf and J. M. Tanur}

\affiliation{Division of Behavioral and Social Sciences and Education,
The National Academy of Sciences and Department of Sociology,
and SUNY Stony Brook}

\address[a]{Miron L. Straf is Deputy Executive Director for Special Projects,
Division of Behavioral and Social Sciences and Education, The National
Academy of Sciences, 500 Fifth St. N.W., Washington DC 20001, USA \printead{e1}.}
\address[b]{Judith M. Tanur is Distinguished Teaching Professor
Emerita, Department of Sociology, State University of New York Stony
Brook, PO Box 280, Montauk, New York 11954, USA \printead{e2}.}

\end{aug}

% ABSTRACT
%
\begin{abstract}
Stephen E. Fienberg is Maurice Falk University Professor of Statistics
and Social Science at Carnegie Mellon University, with appointments in
the Department of Statistics, the Machine Learning Department and the
Heinz College. He is the Carnegie Mellon co-director of the Living
Analytics Research Centre, a joint center between Carnegie Mellon
University and Singapore Management University. Fienberg received his
hon. B.Sc. in Mathematics and Statistics from the University of Toronto
(1964), and his A.M. (1965) and Ph.D. (1968) degrees in Statistics at
Harvard University. He has served as Dean of the College of Humanities
and Social Sciences at Carnegie Mellon and as Vice President for
Academic Affairs at York University in Toronto, Canada, as well as on
the faculties of the University of Chicago and the University of
Minnesota. He was founding co-editor of \textit{Chance} and served as the
Coordinating and Applications Editor of the \textit{Journal of the
American Statistical Association}. He is one of the founding editors
of the \textit{Annals of Applied Statistics}, co-founder and
editor-in-chief of the new online \textit{Journal of Privacy and
Confidentiality} and founding editor of the new \textit{Annual Review of
Statistics and its Application.} He has been Vice President of the
American Statistical Association and President of the Institute of
Mathematical Statistics and the International Society for Bayesian
Analysis. His research includes the development of statistical
methods, especially tools for categorical data analysis and the
analysis of network data, algebraic statistics, causal inference,
statistics and the law, machine learning and the history of statistics.
His work on confidentiality and disclosure limitation addresses issues
related to respondent privacy in both surveys and censuses and
especially to categorical data analysis. He is the author or editor of
over 20 books and 400 papers and related publications. His 1975 book
on categorical data analysis with Bishop and Holland, \textit{Discrete
Multivariate Analysis}: \textit{Theory and Practice}, and his 1980 book on
\textit{The Analysis of Cross-Classified Categorical Data} are both citation
classics. He served two terms as Chair of the Committee on National
Statistics at the National Research Council (NRC) and is currently
co-chair of the NAS-NRC Report Review Committee. He is a member of
the U.S. National Academy of Sciences, and a fellow of the Royal
Society of Canada, the American Academy of Arts and Sciences, and the
American Academy of Political and Social Science, as well as a fellow
of the American Association for the Advancement of Science, the
American Statistical Association, the Institute of Mathematical
Statistics, and an elected member of the International Statistical
Institute.

The following conversation is based in part on a transcript of a 2009
interview funded by Pfizer Global Research-Connecticut, the American
Statistical Association and the Department of Statistics at the
University of Connecticut-Storrs as part of the ``Conversations with
Distinguished Statisticians in Memory of Professor Harry O. Posten.''
\end{abstract}

\end{frontmatter}

\noinbf{MS:} So, Steve, how is it that you came to become a statistician?

\noinbf{SF:} It's actually a long story, because when I was in high
school and entering university, I didn't even know that there was such
a field. I was good at mathematics and I went to the University of
Toronto, which was in my hometown---that's where the best students went
if they could get in. I enrolled in a course called
\textit{Mathematics}, \textit{Physics}, \textit{and Chemistry}. It was
one of the elite courses at U of T, and during the first year, as I
went through my chemistry labs, I never succeeded in getting the right
result when I mixed the chemicals up in the beakers; I realized
chemistry wasn't for me, and so the second year I did only math and
physics. Then there were the physics labs, and I could never quite get
the apparatus to work properly to get what I knew was the correct
answer. I still got an $A$ in the physics lab, because I could start
with the result and work backward and figure out what the settings were
and things like that; but it was clear to me that physics wasn't for me
as a consequence. So that left me with mathematics, and it was in the
%
%f1 #&#
%
\begin{figure}

\includegraphics{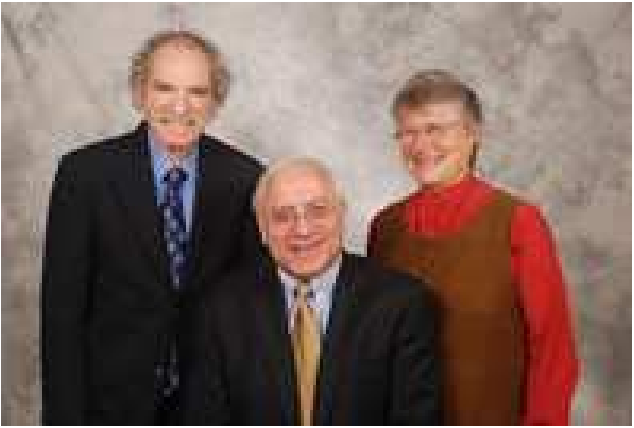}

\caption{Miron Straf, Steve Fienberg and Judy Tanur at the University
of Connecticut, October, 2009.}
\end{figure}
second year that we had a course in probability. So I was being gently
introduced to statistical ideas. Then in my third year there was a
course in statistics that was taught by Don Fraser, and he was
terrific. His course was a revelation, because I didn't know anything
about statistics coming in. Don followed the material in his
\textit{Introduction to Statistics} book and he began with probability
theory and he brought into play geometric thinking throughout. When he
got to inference, it was like magic. Of course, in those days Don did
what was called ``fiducial inference''---he called it ``invariance
theory'' and later ``structural inference''---where you went suddenly
from probability statements about potential observables given
parameters to probability statements about the data. I recall the old
%
%f2 #&#
%
\begin{figure}

\includegraphics{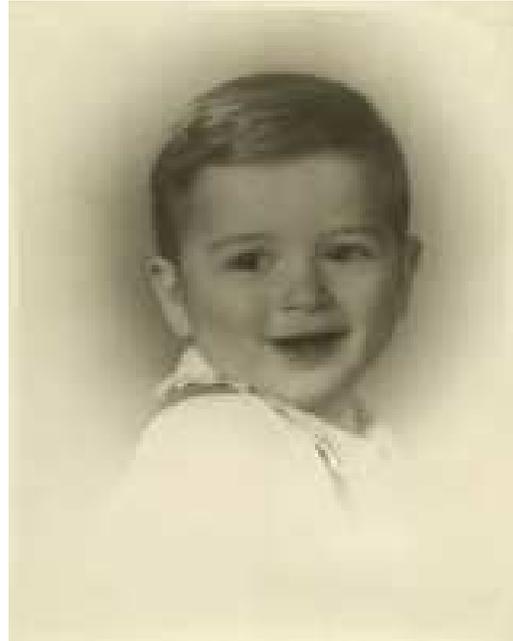}

\caption{Steve as a Toddler in 1940s in Toronto.}
\end{figure}
cartoon by Sydney Harris that people like to reproduce of the two
scientists pointing to a blackboard full of equations, and one of them
points to an equal sign and says, ``And a miracle suddenly occurs
here.'' That's sort of what happened in Don's class. He was a great
lecturer, he was friendly with the students, and it was very clear that
statistics was a really neat thing to do. Thus, in my fourth year I
took three classes involving statistics and probability and then
applied to graduate school in statistics. The rest, as they say, is
history.

%f3 #&#
%
\begin{figure}

\includegraphics{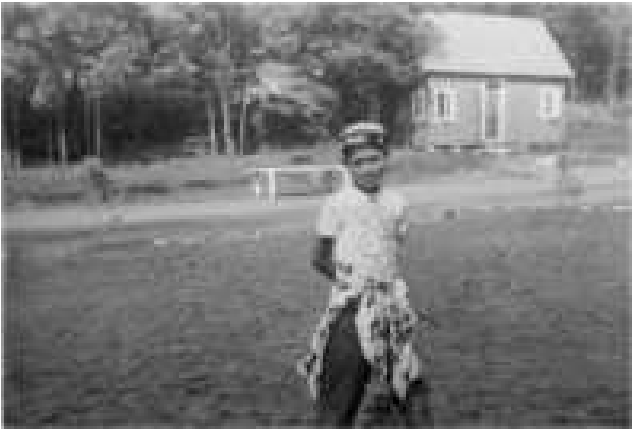}

\caption{Steve at Camp Tamarack, near Bracebridge, Ontario in 1952.}
\end{figure}

\noinbf{MS:} So it was mathematics by elimination and statistics by revelation.
Let's go back a bit. When did you discover that you had an aptitude for
mathematics and statistics? In elementary school? Or high school?

\noinbf{SF:} Not at all. In those days statistics never showed its face in
the K-12 curriculum---this was before \textit{Continental
Classroom}.\footnote{\textit{Continental Classroom} was a series of
television ``course''
broadcasts by NBC on a variety of college-level
topics in the early 1960s. Fred Mosteller taught the course on
Probability and Statistics during 1960--1961} Actually it was K-13 in
Toronto where I was born and
raised. They got rid of grade 13 only decades after I~was in school.
At any rate, although my mother thought I was genius---don't all
mothers think that about their children---I~don't have any memory of
being anything other than just a good student. I was very good at what
passed for mathematics, but even through high\break school I~don't think
I~was truly exceptional, and, besides, we did pretty elementary
stuff---algebra, Euclidean geometry, and then in grade 13 we had
trigo\-nometry. As I reflect on those days, I was good at mathematics, but
certainly not precocious and I only took standard high school math and
with a heavy component of rote and repetition. By the time I got
to grade 13 I was at the top of my class, however, and in the
province-wide exams at the end of the year I was No. 2 in my school.
But I also played oboe in the orchestra and band, and drums in the
marching band, as well as participating in several other extra-curricular
activities. So math wasn't much of a preoccupation and I didn't
know what statistics and probability were all about at all.

\noinbf{JT:} So that explains your broad early work in math,
physics and chemistry as a kind of omnibus course rather than going
directly into math or statistics. So after your undergraduate work
at the University of Toronto, you applied to graduate school; where
did you apply and where did you end up going?

%f4 #&#
%
\begin{figure}\vspace*{7pt}

\includegraphics{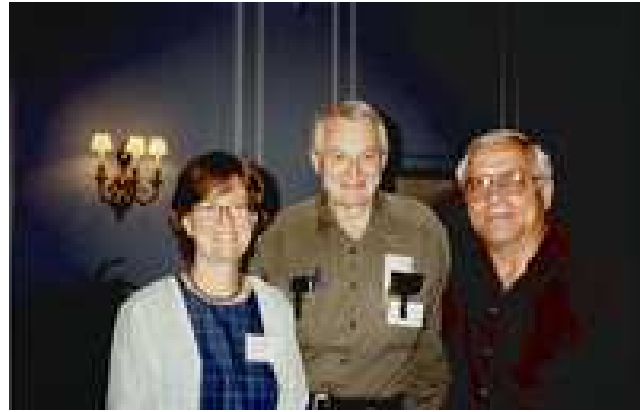}

\caption{Steve with Don Fraser and Nancy Reid at a conference on the
occasion of Don's 75th birthday, June 2000.}
\end{figure}

\noinbf{SF:} Well, at the University of Toronto there had actually been
many people to go into Statistics
from MP\&C. Don Fraser was perhaps the first, but then there were Ralph
Wormleighton, Art Dempster and David Brillinger---they all went, by
the way, to Princeton. The year before me there was John Chambers,
and John had gone to Harvard. I knew John pretty well, and I asked him how
it was at Harvard. He seemed pleased with what he was doing and
I did apply to Harvard and was admitted. I also applied to
Princeton, and in their wisdom they didn't think that I should carry
on the tradition from the University of Toronto, and that made the
decision easier for me.

\noinbf{MS:} Were you disappointed about not being admitted to
Princeton?

%f5 #&#
%
\begin{figure}%[b]

\includegraphics{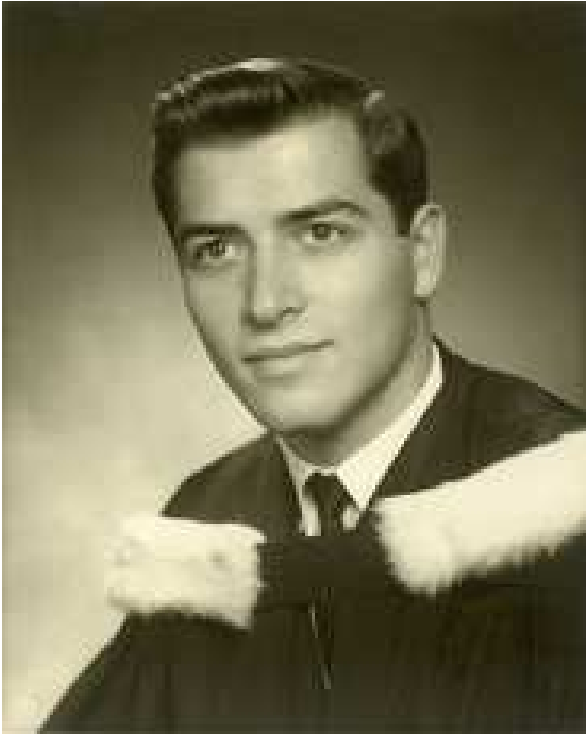}

\caption{Graduation portrait from the University of Toronto, 1964.}
\end{figure}

\noinbf{SF:} Clearly at the time I was. This was my first
rejection, and it prepared me in a way for what was to come when I
submitted papers for publication to major journals! But Sam Wilks,
who was the key person at Princeton with whom I had hoped to work, died
in the Spring of 1964, before I would have arrived.

\noinbf{JT:} By the time you went to Harvard you were already
married, is that right?

\noinbf{SF:} No, I had met my wife Joyce at the University of
Toronto when we were both undergraduates. I was actually working in
the fall of 1963 in the registrar's office, and on the first day
the office opened to enroll people, Joyce came through. And one of
the benefits about working in the registrar's office, besides
earning some spending money, was meeting all these beautiful
women students passing through. That first day I~made a note to
ask Joyce out on a date. The next day she came
through again, this time bringing through another young woman who
turned out to be the daughter of friends of her parents. And I
thought this was a little suspicious, but auspicious in the sense that
maybe I~would succeed in getting a date when I asked her. And the next
day, she came through
again! This time with her cousin! Then I knew that this was really
going to work out. And it did. We got engaged at the end of the summer
of 1964
after I graduated, but we weren't married when I went away to graduate
school. In fact, yesterday I was talking to one of the students at
the University of Connecticut who was a little concerned about
graduate school; it was wearing her down, and I~told her I almost left
after the first semester because I wasn't sure if I was going to
make a go of it, in part because I was lonely. But I did survive, and
Joyce came at the end of the first year; we got married right after
classes ended, and we've been together ever since.

\noinbf{MS:} And where were your children born?

%f6 #&#
%
\begin{figure}

\includegraphics{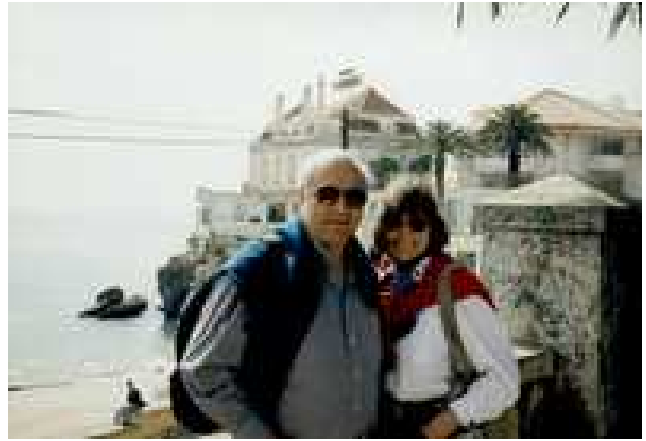}

\caption{Joyce and Steve in Portugal for a conference on privacy and
confidentiality, 1998.}\vspace*{3pt}
\end{figure}

\noinbf{SF:} Ah, \textit{conceived} in various places, born in others.
We believe
that Anthony, my older son, was actually conceived in Scotland, on the
vacation we took just
after I graduated from Harvard. He was born in Chicago, where I had my
first academic appointment, and, indeed, as we traveled across the
country, from Boston to Chicago, Joyce began experiencing morning
sickness (all day long), which didn't make for such a great
trip. Then Howard was born in Minnesota just after we had moved
there and I had joined the University of Minnesota faculty.

\noinbf{JT:} Tell us more about what happened when you first arrived at Harvard.

\noinbf{SF:} Well, one of the reasons I went to Harvard is that they
not only gave me a fellowship, but also a research assistantship to
work with Fred Mosteller. The day after I arrived, I went into the
department because I didn't quite know what a research assistant did,
and I went to see Fred (at the time he was Professor Mosteller, of
course---I didn't learn to call him Fred until later). Fred was busy,
but his assistant, Cleo Youtz, said he would like to have lunch with
me. So I came back for lunch, and we went to the Harvard Faculty Club.
Fred was being very courteous, and he suggested I order the horse
steak, a special item on the faculty club menu at the time. And the
horse steak came---I'm not sure if you've had horse steak---it's not
\textit{quite} like the kinds of steaks we normally order, it's a
\textit{little} bit tougher. I cut my first piece of horse steak, I put
it in my mouth and started to chew. And then Fred began to describe
this problem to me. It was about assessing probability assessors.
I~didn't understand a thing, and he's talking away, and I'm chewing
away. Then Fred asked me a question, and I'm chewing away. At this
point, he pulled an envelope out of his pocket and on the back of it
there were these scribbles. He handed it to me, and I'm \textit{still}
chewing because you really can't eat horse steak except in very small
bites. It turned out that the scribbles were notes from John Tukey
about this problem. In fact, this was a problem that John and Fred were
working on for some larger project, and my job was to translate the
chicken-scratches on the back of the envelope into something
intelligible, when I didn't know anything about what was going on. I
worked at it for a while, and then Fred slowly told me what John's
jottings meant, and the key idea was that for assessing probability
forecasts, you have to look not just at the equivalent of means, or the
bias in them (known technically as calibration), but also at the
equivalent of variability (how spread out the forecasts are). Actually,
that was a very important lesson, although I didn't have any clue about
it in my first months at Harvard.

Over the course of my first fall at Harvard, I discovered a
paperback book called \textit{The Scientist Speculates}: \textit{An Anthology of
Partially Baked Ideas}, edited by Jack Good, with whose work I later
became very familiar. In it was a short essay by Bruno de Finetti
on assessing probability assessors, and de Finetti's ideas went into
the technical report I wrote up on the topic with Fred and John.
Fifteen years later, at the Valencia I Bayesian meeting, Morrie
DeGroot and I began to work on the problem and ultimately wrote
three papers on the topic of calibration and refinement of probability
forecasters, heavily influenced by that first research exercise
with Fred.

%f7 #&#
%
\begin{figure}

\includegraphics{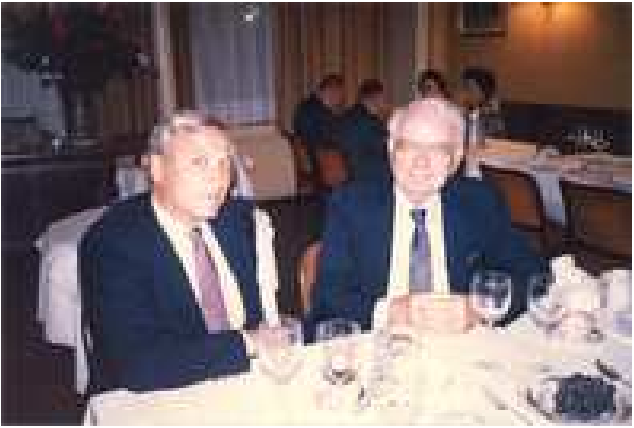}

\caption{Steve dining with Fred Mosteller at ISI meetings in Paris, 1989.}
\end{figure}

\noinbf{MS:} I wanted you to talk about Fred. Fred has been
a very influential person in your career, and not just during your
thesis. Maybe you want to tell us a little bit more about how he
influenced your lifek and also how you came to go from Harvard to
Chicago.

\noinbf{SF:} Well, during that first year I worked on several
problems with Fred and I wrote up some memos, but they never quite
moved into papers at the time. Fred was pretty busy, and I got
interested in Bayesian inference and multivariate analysis. I had
begun to take an interest in Bayesian
methods, having participated as a first year student in a seminar
across the river at the
business school run by Howard Raiffa and Bob Schlaiffer. At
the time, Art Dempster was the person who seemed to be most involved
in these Bayesian things and multivariate analysis, so I began to meet
with him. In the process of working
with Art, I met George Tiao, who was visiting the Business School with
George Box for the
year. As a consequence, George and I wrote a paper together on
Bayesian estimation of
latent roots and vectors but it just didn't look like it was going to
be a
thesis problem.

The next summer, Fred ran into me in the hall and
said he had some problems that I might like to work on. Fred had
become deeply involved in the National Halothane Study at the NRC
and, unlike most NRC studies, he and others---Tukey, John Gilbert,
Lincoln Moses, Yvonne Bishop, to name a few---were actually
analyzing data and creating new methods as they went along. The
data essentially formed a giant contingency table and Fred got me
working on a few different problems that ultimately came together as
the core of my thesis. In the process I~collaborated on separate
aspects of the work with John Gilbert, Yvonne Bishop and Paul
Holland. I~did most of the work in 1967 and that was the summer of
``The Impossible Dream,'' when the Boston Red Sox won the pennant. I would
work into the wee hours and go to Fenway Park and sit in the
bleachers for the afternoon games. Professional sports where cheap
in those days. We also used to go to Boston Gardens for Bruins and
Celtics games. Fred was also a Red Sox fan and he actually got
tickets for some of the 1967 World Series games. I~was envious, but
when I returned to Boston in 1975 on sabbatical we both were able to
get World Series tickets.
I got tickets for game 6 and Fred got them for game 7!

Fred introduced me to lots of other statistical problems. I was also
his TA
one year, working with Fred and Kim Romney who was in the Social
Relations department at the time. Then the time came to get a job,
and Fred said to me, ``Where would you like to go?'' Things were
different in those days, as you will recall from your days at
Chicago. We went through the list of the best places in the field,
at every one of which Fred had a friend. He called up John Tukey
at Princeton, he called up Erich Lehmann at Berkeley, Lincoln Moses
at Stanford and Bill Kruskal at the University of Chicago. I
either got offers without showing up for different kinds of jobs at
these places or
I got invited out for an interview. When I was invited to interview at the
University of Chicago, it just seemed like a really neat place. All
the faculty members were friendly. The temperature in January was
really cold,
but I liked everything about the university from the people to the
architecture; it looked like a university.
Leo Goodman was there on the faculty and he had done work that was directly
tied to contingency table topics in my thesis. Chicago just seemed
like a great place
to go to, so I~did.

\noinbf{JT:} It was there that you first met Bill Kruskal and
started being influenced by him?

\noinbf{SF:} Bill Kruskal was the department chair at the
time, and I barely got in the door before he began talking to me about
a slew of different statistical problems\ldots

\noinbf{JT:} Without horse steak?

\noinbf{SF:} Yes, without horse steak. Bill would just come
and say, ``What do you know about this?'' And one of the first topics we
actually discussed was political polls. This was the summer of 1968;
there was a lot going on politically in the U.S., and the \textit{Sun Times
Straw Poll} was showing up in the newspaper regularly. Two of the
key questions were: What was their real methodology? How accurate were
their predictions? I began to save the data from the newspaper
reports and work on the question of variability and accuracy. Then
Bill got me to do a trio of television programs with Ken Prewitt and
Norman Bradburn on a special series that aired at 6 o'clock in the
morning when nobody ever watched. But right from the beginning, Bill
and I interacted; he introduced me to Hans Zeisel in the law school,
to people in the business school, in sociology. It was really hard to
trail after Bill, because he was interested in everything in the
university and outside, and almost everything we discussed seemed
pretty neat. So, as I~launched my professional career at Chicago, I
tried to do something similar---not precisely the same as the way Bill
did things---but similar.

\noinbf{MS:} Bill was a real Renaissance man, and I presume you
were a recipient of his many clippings from newspapers.

\noinbf{SF:} Well, the clippings started when I was in my
first year---he's the one that started to give me the \textit{Sun Times
Straw Poll} clippings. But it wasn't just clippings. Bill would leave
library books for me in my box; he would go to the library, which
was on the second floor of Eckhart Hall, the building we were in,
and he would browse---people don't do that today---the stacks are
closed. He would come back, armed with books, and he would share
them with his colleagues and get Xeroxes of pages. And this
continued up through the 1980s. I~would always get packets of
different materials from
Bill, including copies of letters to somebody else that would say:
``I~hope you don't mind my sharing this with a few of my closest
friends and colleagues.'' I had this image that he was making
hundreds of Xeroxes to send around the world.

\noinbf{MS:} And before that, carbon paper. So, tell us a bit
about your life after Chicago.

\noinbf{SF:} The University of Chicago really was a great place for me
to work. I
had a second appointment in theoretical biology, which was
interesting because I~had never taken a course in biology as a student.
And actually it
was a very formative experience, because it taught me that I could go
into an area that I~had never studied, never learned anything about,
and learn enough for me to make a difference in the application of
statistics. I wrote
papers on neural modeling, and I~wrote papers on ecology; I didn't
do a lot of genetics, but I read genetics papers and books because I
included that material in the course on stochastic processes that I
taught. Unfortunately, Chicago wasn't the safest of places in those
days, and Joyce made it pretty clear that she wanted to live in a
place where our children could play in the backyard by themselves, not
under adult supervision 100 percent of the time. So I began to be
receptive to conversations with people from the outside, and soon I
was approached by one of my former students, Kinley Larntz, who had
just joined the University of Minnesota. They were looking for a
chair for the newly created Department of Applied Statistics, as part
of a School of Statistics. So after four years at Chicago, I became
an administrator as well as researcher and teacher.

%f8 #&#
%
\begin{figure*}

\includegraphics{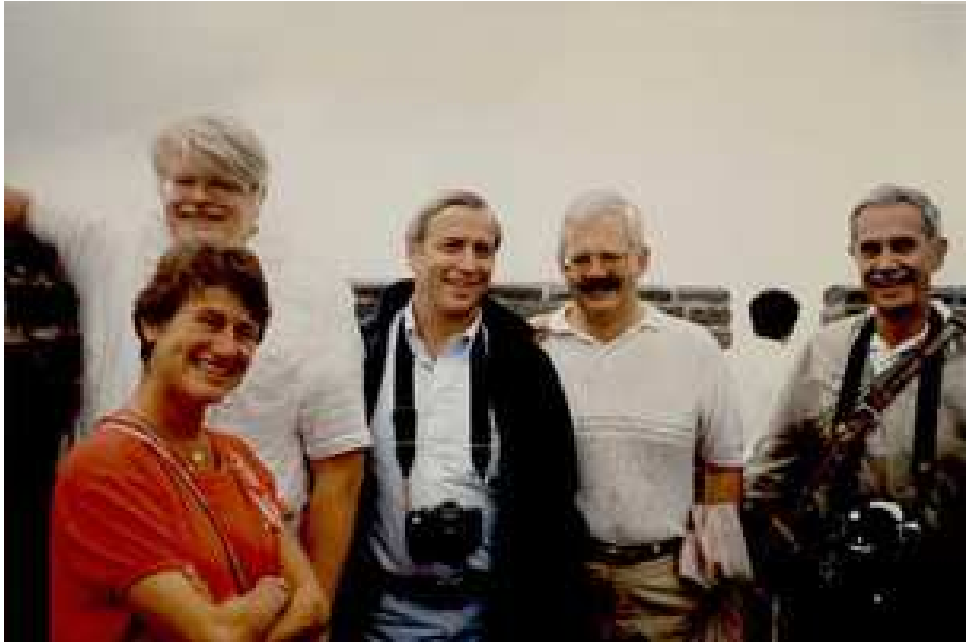}

\caption{Judy Tanur, John Bailar, Steve, Henry Block and Jim Press at a
conference in Bejing, 1987.}
\end{figure*}

\noinbf{MS:} Did you work with Seymour Geisser there?

\noinbf{SF:} The School of Statistics was an interesting
idea. Minnesota had had a statistics department, and it had run into
some problems over the years. The university came up with this plan
to reinvigorate statistics, and they created the School of
Statistics. Seymour was the director, and the School was supposed to
have three departments. There was the old statistics department,
renamed as the Department of Theoretical Statistics, there was the new
applied department that I was chairing, and there was the Biometry
department in the School of Public Health. But the biometry faculty
didn't really seem to want any part in this, and so they resisted, and
ultimately the school had two departments plus the Statistical
Center---the consulting center that was associated with our department
on the St. Paul part of the Twin Cities campus. Seymour and I interacted
throughout my eight years at Minnesota, but we never wrote a paper together.

\noinbf{JT:} I want to take you back a little more. You
talked about these two giant figures who were colleagues and
mentors---Fred Mosteller and Bill Krus\-kal. How do you see how they
shaped your career, your interests---not only technical, but
practical?

\noinbf{SF:} One of the things I didn't know as a graduate
student was how easy it would be to work on and contribute to new
problems and new areas of application. The worst fear of a graduate
student---well, the worst fear
is that they won't finish their thesis---the second fear is they won't
have a new idea, and, in fact, 80\% of students never publish anything
other than their thesis. But Fred was going from area to area: when I
arrived at Harvard he had just published \textit{The Federalist Papers}
with David Wallace; while I was there he was leading the effort on the
Halothane report; I worked with him evaluating television rating
surveys from Nielsen and other companies for a national network (that
was a consulting problem). He just seemed to work around the clock on
all sorts of different topics, and so I~figured that's just what a
statistician did. It's funny because, in some senses, clearly,
everyone didn't behave like Fred, as we all know. But that was my
model! So when I
got to Chicago and Bill acted in the same way, and Paul Meier in
addition, that seemed like a natural way for me to do work as a
statistician. They
seemed to work around the clock on statistics, so I did too.

Now Fred liked art; in later years he actually took up
reproducing art and it showed up in his office. When I~was a
graduate student I went into his office one day and there was a
picture by Escher, the Dutch artist, called ``The Waterfall'' and I
was very surprised because I had been introduced to Escher as an
undergraduate. Escher's work showed up on the cover of a book called,
\textit{Introduction to Geometry}, written by Donald Coxeter---the great
geometer at the University of Toronto. I had three courses on different
aspects of geometry from
Coxeter. This influenced some of my thesis research---and I still do
some geometry---but I also learned about Escher from Coxeter! And
there was this Escher print in Fred's office which I recognized immediately.
Fred told me where he had purchased
it, and shortly afterwards I went off to the store. I still own two
Escher prints as a
consequence, ones that I couldn't afford to buy today, all because of Fred.
Fred and I~would occasionally go off to museums, and while we looked at
the art we
would talk about statistics, art and other topics.

%f9 #&#
%
\begin{figure}

\includegraphics{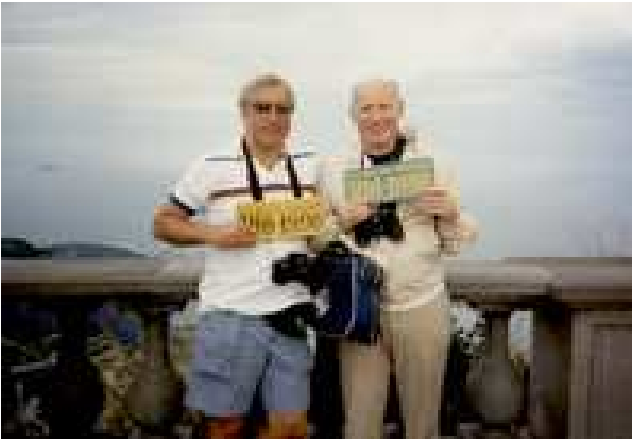}

\caption{Steve and Seymour Geisser, attending a Bayesian Workshop in
Rio de Janeiro, Brazil, summer 1990.}\vspace*{3pt}
\end{figure}

Both Fred and Bill were Renaissance men and I~didn't know how I would
do things in the same way they did, but it became very clear to me that
just doing papers in the \textit{Annals} and in \textit {JASA} wasn't
enough. While I had colleagues whose careers looked like that,
I thought I should be doing something different with my career. I was
easily seduced into all these other activities---and everything was so
much fun. For example, Dudley Duncan, the sociologist, called me one
day and asked me if I would join an advisory committee set up by the
Social Science Research Council on social indicators in Washington.
I~hadn't been to Washington since I was 7 years old and I went off to
this meeting and then spent eight years interacting with giants in the
field of sociology and survey methods! That experience just reinforced
the way I was using my statistical knowledge in diverse applications.

And of course Bill and Fred would just sort of nudge me once
in a while to get things done that they cared about deeply. In
particular, Fred wanted to see the log-linear model work that his
students had done for the Halothane study appear in a book. Fred was
big on books. And as I left Harvard, he gathered together all the
different students who had worked on different aspects of
contingency table anal\-ysis---Yvonne Bishop, Dick Light, myself and Paul
Holland, who was a
junior faculty member, for a meeting at his house. There were also a
couple of other faculty members who sort of disappeared by the wayside
in this enterprise, there were a few more graduate students---Gudmund
Iversen who ended up at Swarthmore, for example---and Fred said, ``We
need to have a book on this.''

But we didn't have Fred's grand picture in mind and the book didn't
begin to take shape until long after I~had joined the faculty at the
University of Chicago. I~taught a contingency table course in my first
year there and it included the first three Ph.D. students I~worked
with---Tar (Tim) Chen, Shelby Haberman and Kinley Larntz. Shelby
extended\break Yvonne's code for multi-way tables and this inspired his
thesis. I began to use iterative proportional fitting on new problems
and this triggered a paper on multi-way incomplete tables and a draft
of the first book chapter. But then everything progressed rather
slowly, and the book took a full six years to produce. Fred kept
pushing the book behind the scenes.

One of the things I learned is the time to produce a book
goes up as the power of the number of authors. It would have taken
less time if I had written the book myself instead of with Yvonne and
Paul. But while we worked at the core of the enterprise, the three of
us had
different conceptions of some materials, and this slowed us down.
Fred was a full partner, pushing us to ``get the job done.'' He edited draft
chapters over and over again, and Dick Light contributed big chunks to
the chapter on measures of association, which Paul and I redid and
integrated with the asymptotics chapter. If everyone who had come to
Fred's house back in 1968 had become involved, we might still be
working on the book today! Fred didn't want his name on the cover of
the book. So we had this back-and-forth. The book ended up with five
names on
the title page; it's Yvonne Bishop, Stephen Fienberg, Paul
Holland, with the collaboration of Frederick Mosteller and Dick Light;
Dick had contributed to a chapter in the book and Fred had
contributed to the whole enterprise.

\noinbf{JT:} The book, which many have called the ``Jolly Green
Giant'' because of its cover, really put you on the map. In fact,
that's how we met, when I took the short course the three of you gave
based on the book in 1976 at the Joint Statistical Meetings.

\noinbf{SF:} We actually met earlier, when Fred organized a
meeting in Cambridge to discuss the ASA-NCTM book projects that
ultimately produced \textit{Statistics by Example} and
\textit{Statistics}: \textit{A
Guide to the Unknown}, your first \textit{magnum opus}. I was a bit
intimidated since you seemed to be the organizer for \textit
{Statistics}: \textit{A Guide to the Unknown}, and so we just didn't talk
much.

\noinbf{MS:} Steve and I met around the same time as well. I
remember his coming to Chicago to interview and talking about the
geometry of $2\times2$ tables. I asked him a question which he didn't really
answer and then he wrote a paper about that problem several years
later!

\noinbf{SF:} But when I got to Chicago you were one of the few
good students who didn't take my contingency table course. You were
too busy campaigning for Hubert Humphrey and worrying about weak
convergence!

\noinbf{MS:} Well, one of the things that you have advanced in
that book and elsewhere derives from the geometric structure that gave
you so much insight into what's going on in these tables. Now, you
mentioned taking geometry at Toronto, and we know R. A. Fisher was
influenced by this, so how did that play out in the later research?

\noinbf{SF:} It's come into play in an amazing sort of way.
If you look at the cover of \textit{Discrete Multivariate Analysis},
there is an artist's depiction of the surface of independence for a
\mbox{$2\times2$} table. You'd hardly know it was a hyperbolic paraboloid sitting
inside a tetrahedron by the time the artist got done with it, and you
see one dimension of rulings---a hyperbolic paraboloid has two
dimensions of essentially orthogonal rulings---and those are things I
actually learned from Coxeter in that course on the Introduction to
Geometry. And so my first work actually drew upon that; I wrote a
paper with John Gilbert on the geometry of $2\times2$ tables that appeared
in \textit{JASA} and published a generalization in the \textit{Annals}, and
I always thought about contingency tables and other statistical
objects geometrically. Don Fraser thought geometrically, and so
you're always up here ``waving arms'' in some abstract space, and he would
always wave with his arms. And I think in high-dimensional space in
some sense, although obviously we don't see in high-dimensional space.
But a lot of statistics is projecting down into lower-dimensional
spaces. I~had left the geometry stuff behind, except for
motivation, until I got into confidentiality research in the 1990s.

In the 1990s there was a paper, unpublished for five years by Persi Diaconis
and Bernd Sturmfels. Persi was at Cornell and Bernd had been at
Cornell but moved to Berkeley.\vadjust{\goodbreak} In the paper, they talked about the
algebraic geometry structure associated with contingency tables. This
turned out to be right at the heart of what I needed for my problem,
and so I learned algebraic geometry, which I had not really studied
carefully before. I learned at least enough to bring my problems to
Bernd for help. And one of the things I realized is that figure on
the cover of Bishop, Fienberg and Holland was being used by
algebraic geometers in a different context; it's called a Segre
Variety, named after Corrado Segre who was one of the fathers of
algebraic geometry. That work is now reflected in
the theses of a couple of my former Ph.D. students and lies at the heart
of a lot of what I've been doing over the last several years,
including recent work on algebraic statistics and network models.

%f10 #&#
%
\begin{figure}

\includegraphics{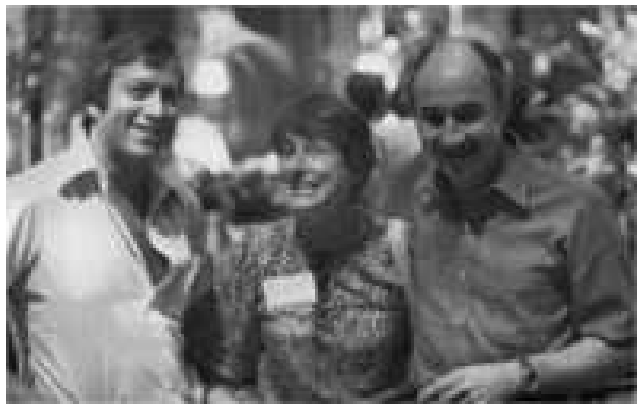}

\caption{Steve, Judy Tanur and Morrie DeGroot, Joint Statistics
Meetings, 1978.}
\end{figure}

\noinbf{JT:} I think I derailed you sometime back where you
were talking about the trajectory of your career. And we've left you
at Minnesota. Can you tell us why you left?

\noinbf{SF:} Minnesota was a giant bureaucracy. It was a big,
big university, and one of the moments that convinced me of
this was after I had presented a report, prepared with colleagues from
around the university, to the president and the vice-presidents on the
teachings of statistics at the university, where I had pointed out
that 40~different departments or units where teaching statistics or
courses in which statistics represented a serious part of the
activity. Virtually all of this was going on with little or no
coordination with the School of Statistics. And then I met him [the
president of the university] about a month later at a reception.
Joyce and I~were going through the reception line, and I shook his
hand, and he asked what department I was from. I said applied\vadjust{\goodbreak}
statistics, and he said, ``\textit{Do we have a statistics department at
the University of Minnesota?}'' At that point I said to myself, ``Oh
my goodness!'' and I~understood where the School of Statistics and
my department stood in the big picture of the university.\looseness=-1

A year or two later, I was wooed by friends at another Big Ten
university, but the right offer didn't quite come to pass. In the
mid-70s I was working as an associate editor for the \textit{Journal of the
American Statistical Association}, initially with Brad Efron as theory
and methods editor, and then with Morrie DeGroot. Later I became
Applications and Coordinating Editor of \textit{JASA}, and so Morrie
and I
worked together. We had become friends a number of years earlier,
drinking in a bar together at an IMS regional meeting. Morrie and Jay
Kadane, who had joined the Department of Statistics at Carnegie Mellon
in the early 70s, and I would interact at the Bayesian meetings that
Arnold Zellner organized twice a year. They both knew that I had
flirted with the possibility of leaving the University of Minnesota,
and they said, ``You should
just come to Carnegie Mellon; you could bring the rest of \textit{JASA} over
and we'd have the whole journal. Besides, it's a great place.'' So
they worked on the possibility of an appointment for me. When I came to
interview, it
wasn't just to meet with the Dean, and with Jay and Morrie and the
people in the department that I knew. They took me to see the
president of Carnegie Mellon (CMU), who at the time was
Richard (Dick) Cyert. Dick was an economist but also a statistician!
He took courses from Hotelling and Cochran at Columbia as a graduate
student, and although his degree was in economics, he always thought
that he was a statistician as well. In particular, he was a member
and Fellow of ASA. Dick helped found the CMU Department of Statistics
in the mid-1960s when he was the dean of the Graduate School of
Industrial Administration. He was actually the acting chair at
the outset until Morrie took over. So the staff ushered me into his
office. I~had never met Dick before, but that afternoon I spent two
hours with the president of Carnegie Mellon. And I told you about my
interaction with
the president of the University of Minnesota! Here I am sitting with
the president of Carnegie Mellon, this great university, and he's
telling me how important it is for me to come to Carnegie Mellon and
what I'm going to do for the field of statistics. He said, ``If you
come here, everything you do will be called statistics. You will get
to change the field.'' So I came. And I hope that I've changed parts
of the field.\vadjust{\goodbreak}

\noinbf{MS:} Cyert was a visionary, and really led the
Graduate School of Industrial Administration to a high place among
business schools and understood that he needed quantitative strength,
and so he influenced you and supported you. I wanted to ask about one
of your greatest honors, and that is your election into the National
Academy of Sciences. Where were you and how did you get the word?

\noinbf{SF:} Most people don't know what goes on at the National Academy---it's
like a secret society---and it's selection process is Byzantine,
running over the
course of one or more years. At the end,
the NAS members meet in Washington at the annual meeting in a business
meeting and they elect the new members. That happens between 8:30 and
9 in the morning; then they take a break in the meeting and everybody
rushes out to find a telephone and they call their friends and the
newly elected members to the section to congratulate them. This was
in the spring of 1999, and I was teaching---actually that year I was
teaching an introductory statistics class, so I had to be there
relatively early---it was just at 9 o'clock, I was opening the door to
my office, and the phone rang. I answered and it was several friends,
mainly demographers---Jane Menken, Doug Massey, a couple of
others---and there was a chorus on the phone saying ``Congratulations,
you've been elected to the National Academy!'' I was floored,
because I'm not quite sure whether they knew, a year
or so earlier I wouldn't have been eligible, because I~was born and
raised in Canada, and I hadn't become an American citizen until
January 1998. Thus being elected the next year was a special
honor.

%f11 #&#
%
\begin{figure}

\includegraphics{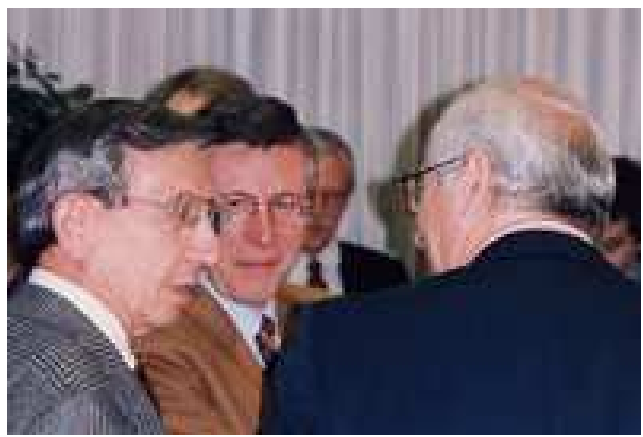}

\caption{Richard Cyert, Dennis Gillings and Steve, at a National
Institute of Statistical Sciences Board of Trustees Meeting, 1993.}
\end{figure}

\noinbf{JT:} You have received many other awards and honors;
that must be very exciting.

\noinbf{SF:} Well I would be lying if I said that receiving
honors and awards is not fun, and each is always very special. But I
am reminded about something that Fred taught me. He said that awards
and honors are really not for the people who get them, but they are
for the field. Of course the person getting the honor benefits, but
the field benefits more, for example, when statisticians get elected
to the National Academy of Sciences. In that sense we don't have
enough big awards.

\noinbf{MS:} There are some of our colleagues who are happy
that there isn't a Nobel Prize in Statistics, and as a consequence
statisticians cooperate more with one another than scientists in other
fields. Do you agree?

\noinbf{SF:} Well, I think if we follow Fred's reasoning we
would all be better off with a Nobel Prize in Statistics because once
a year all of the newspapers and media in the world would focus on our
field and the accomplishments in it. What most statisticians don't
know is that there almost was a Nobel Prize!

The story goes back several decades when Petter Jacob
Bjerve, who was the director of Statistics Norway, began to raise
funds for a Nobel Prize in Statistics. He was off to a good start
when he ran into a political obstacle. Those in charge of the prize
in Economic Sciences objected because, they argued, their prize
encompassed a large amount of what was important in statistics. In
the end Bjerve abandoned his quest, and the money he raised was left
in a special account in Statistics Norway. Finally, the government
auditors forced Statistics Norway to close this account and our
colleagues there decided, among other things, to use the funds to host
a special international seminar, to which they invited statisticians
such as Fred Smith from the UK, Jon Rao from Canada, Wayne Fuller, me
and a few others. They paid for our spouses to come as well and we
got the royal (small R) treatment, with relatively fancy hotel rooms
and outstanding dinners. So in this sense you could say that I ate
the Nobel Prize in Statistics, although there is no public record and
it doesn't show up on my CV.

\noinbf{JT:} You've been active in several committees and
panels and so forth, including at the National Acad\-emies before and
after your election as a member---what stands out particularly from
those?

\noinbf{SF:} Well, of course this is Bill Kruskal at
work---most statisticians who are going to read this interview don't
know the history---Bill Kruskal founded the Committee on National
Statistics (CNSTAT) at the NAS. It was an outgrowth of the 1971 Report
of the President's Commission, chaired by Allen Wallis and co-chaired
by Fred Mosteller; and Bill talked the people at the National
Academies, and the National Research Council (NRC, its operating
wing), into creating a committee although there was no external
funding, and
the NAS really had to put up resources. Bill ultimately got some money
from the Russell Sage Foundation to tide the committee over with a
part-time staffer---Margaret Martin, who was and is absolutely
fabulous and with whom the three of us have worked---and the committee
slowly got going. Bill was succeeded by Con Taeuber. At that time I
actually was on another committee, on the rehabilitation of criminal
offenders, but Miron was working for CNSTAT and I would
run into him on occasion. I got to join CNSTAT a year or so later while
I was
still doing the work on criminal justice. Getting involved in CSTAT
was like all these other activities I have been describing---I was exposed
to lots of
new ideas and problems to work on. I was like a
kid in a candy shop! The committee didn't have a lot of projects
then, but I just got to look around the Academy and the Federal
government, and there were possibilities everywhere. I~could only do
so much, but I pushed the staff to do other things and got my friends
on the committee to lead panels. By the mid-80s the committee was
humming and there were all these neat activities on census
methodology, on cognitive aspects of survey methodology, statistical
assessments as evidence in the courts, sharing research data---there
was just no end.

\noinbf{MS:} I wanted to ask about one of them in particular,
which Judy chaired and which you were instrumental in creating, and
that is
Cognitive Aspects of Survey Methodology. When you were inducted into
the American Academy of Political and Social Sciences, you referred to
that in your speech as one of the most important activities that you
had participated in. Why was this and how did it affect your work?

%f12 #&#
%
\begin{figure*}

\includegraphics{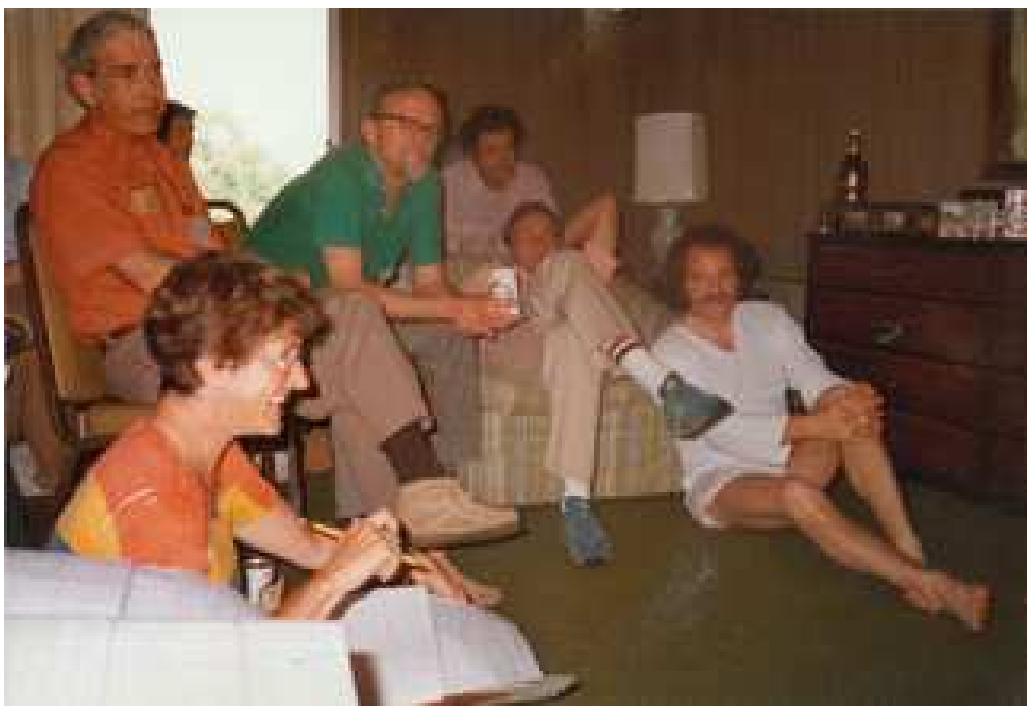}

\caption{Participants at 1983 CNSTAT Workshop on Cognitive Aspects of
Survey Methodology watching a survey interview video, from left to
right: Kent Marquis, Judy Tanur, Phil Converse, Lee Ross, Steve (in
upholstered chair), Miron Straf.}
\end{figure*}

\noinbf{SF:} Well, sample surveys is a very strange part of
statistics. In my department, nobody else really does it, in the
research sense. People think the theory is settled. But \textit{doing}
surveys is \textit{really} hard. The measurement problems are enormous.
Designing questionnaires is a big, big problem. In the 1970s I got
interested in the National Crime Survey on Victimization through the
SSRC committee on social indicators in Washington on which I served.
I learned about the difficulties in counting victimization
events.\vadjust{\goodbreak}
In 1980 Al Biderman, who was involved in the re-design effort for
the victimization survey, brought together a few people from the
re-design project with cognitive psychologists to ask if we could
learn something from cognitive science. I~thought this was just
terrific because I could see ways that I could take methodological
statistical ideas and really intertwine them with the theoretical
ideas that came out of cognitive psychology. As a consequence,\break
I~pushed for that CNSTAT activity even though others thought it made no
sense. I was part of the CNSTAT workshop that you and Judy
organized---Judy and Beth Loftus and I wrote a series of 4 papers on
cognitive aspects of surveys afterward. I was also on the SSRC
council, and we created a committee that followed up on those
activities. It brought in new people to the enterprise, and it helped
get these ideas embedded in the statistical agencies. Janet Norwood
ran with the idea at BLS. It was part of the culture at NCHS at that
time because Monroe Sirkin was at the CNSTAT workshop and a moving
spirit in establishing a cognitive laboratory at NCHS. The Bureau of
the Census was actually the last of the big three agencies to create a
separate laboratory facility---but they did---and the influence spread
because the associated ideas changed research at the boundaries of
survey methods and psychology in a variety of different ways. The
reason I am especially proud of this activity is because you'd hardly
know that there was any statistical theory or methodology lurking
behind it, but there really was.

\noinbf{MS:} It's really had a profound effect on the survey
field, and now in many places it's commonplace---concepts of cognitive
interviewing and all that.

You've been especially close to your students, fostering
them personally as well as professionally. Pictures of you attending
weddings of your students appear frequently on websites in your honor.
So could you tell us a little about your personal interactions with
your students.

\noinbf{SF:} Well, in the early years the students were my
contemporaries. In fact, I had a couple of students who were older than
I was. Kinley Larntz was not only my Ph.D. student and collaborator, but
we were good friends, and remain so. Over the years I got a little
older than my students, and when I moved to Carnegie Mellon I really
had the opportunity to have a different kind of student, and with them
different kinds of interactions. We were a small department in those
days and I~interacted with lots of students, not just those whose
research I~supervised. Each of the students I~worked with then was
interested in a somewhat different topic; they went in different
directions, and we remained close in most instances.

%f13 #&#
%
\begin{figure*}

\includegraphics{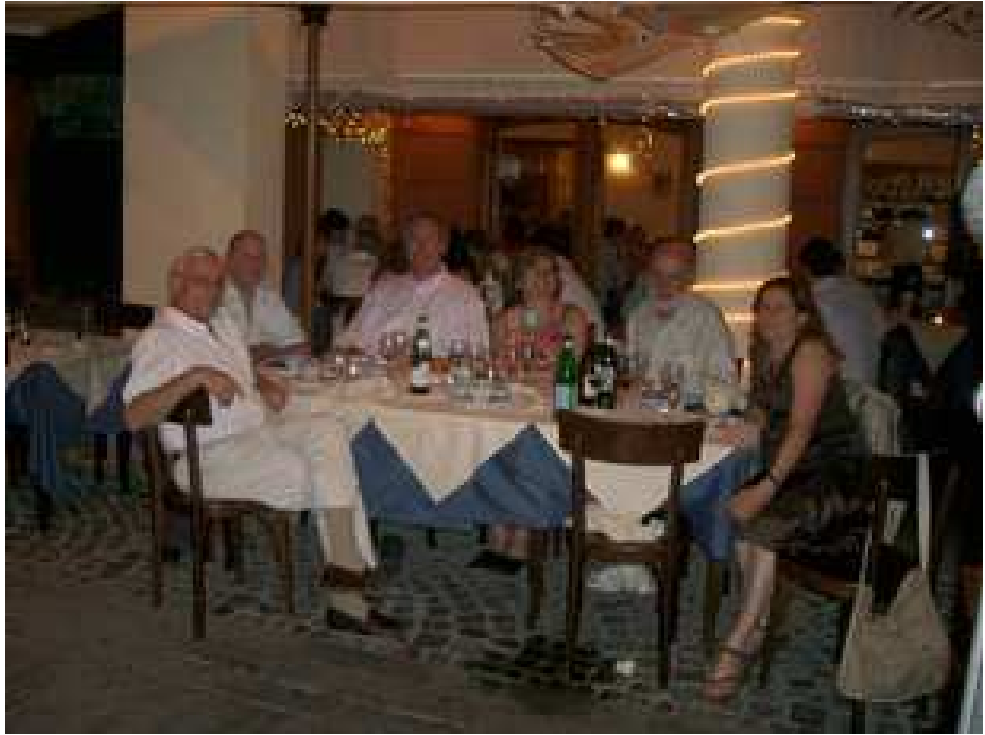}

\caption{Steve with friends at the Objective Bayesian Analysis meeting
in Rome, June, 2007. From left to right:
Steve, Larry Wasserman, Jim Berger, Susie Bayarri, Robert Wolpert, Isa
Verdinelli.}\vspace*{6pt}
\end{figure*}

%f14 #&#
%
\begin{figure*}

\includegraphics{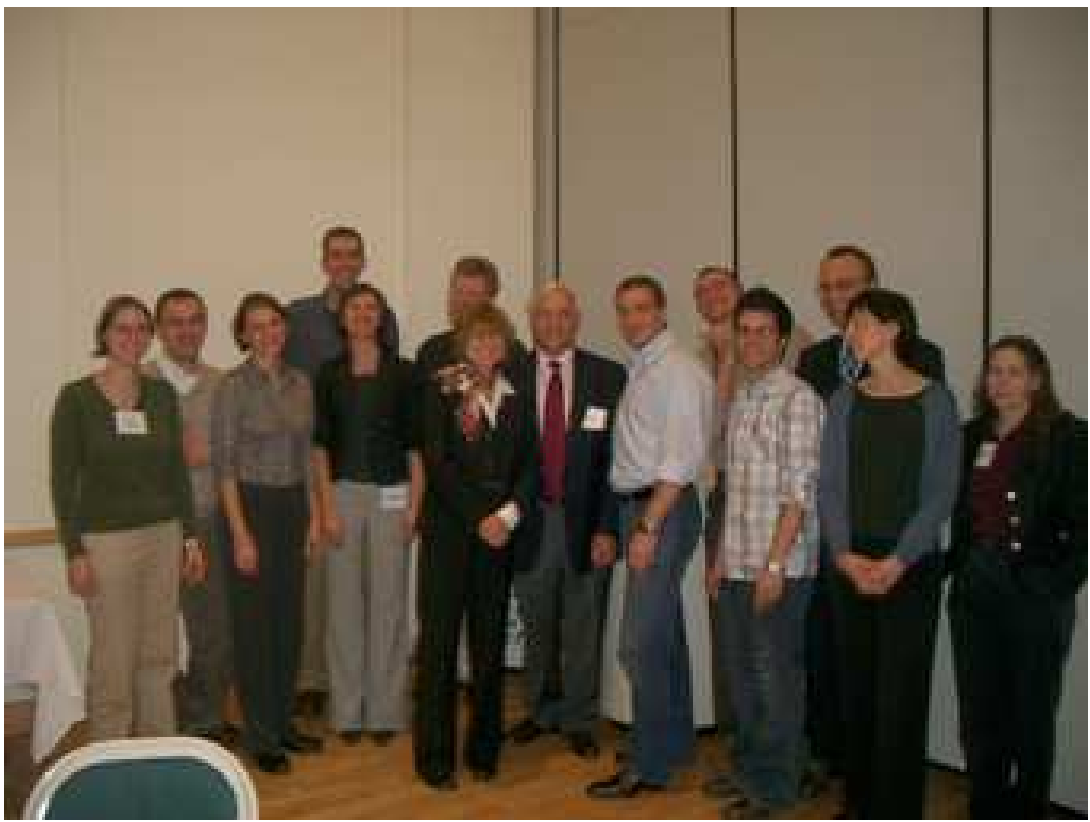}

\caption{Steve with his wife Joyce and many of his former graduate
students at a 65th birthday celebration at Carnegie Mellon, October,
2007. From left to right:
Ellie Kaizer, Edo Airoldi, Elena Erosheva, Jason Connor, Sesa Slavkovi\'
c, Mike Meyer, Joyce, Steve, Alessandro Rinaldo, Justin Gross, Russ
Steele, Adrian Dobra, Amelia Haviland, Elizabeth Stasny.}\vspace*{-3pt}
\end{figure*}

%f15 #&#
%
\begin{figure*}

\includegraphics{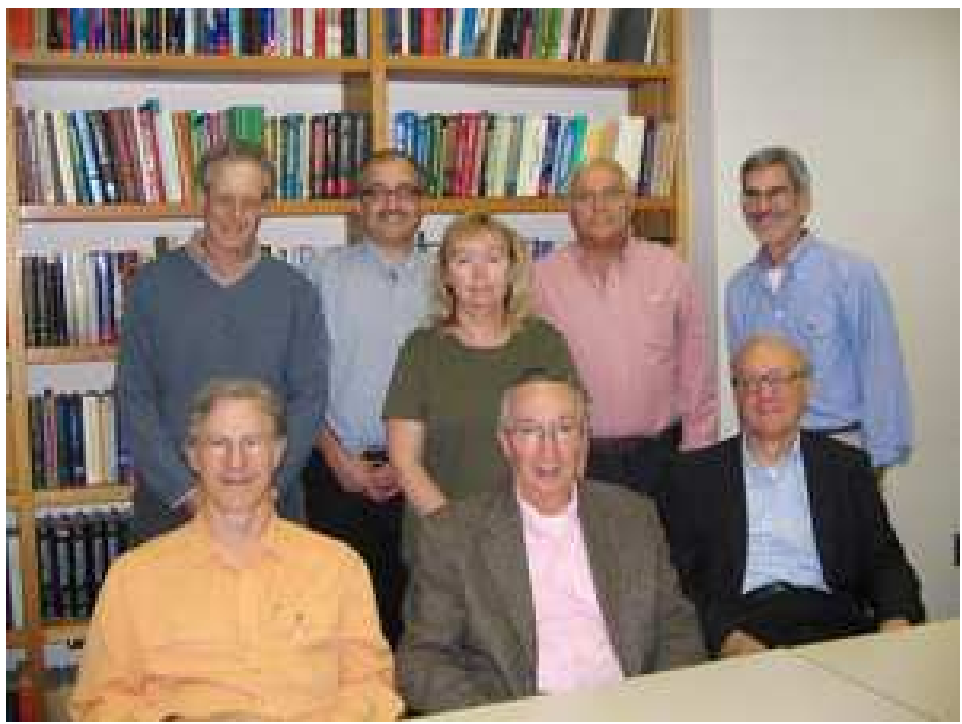}

\caption{The longtime members of the Carnegie Mellon Department of
Statistics in the DeGroot Library, 2011. Back row: Rob Kass, Mark
Schervish, Steve, Joel Greenhouse; middle: Margie Smykla; bottom row:
Jay Kadane, Bill Eddy, John Lehoczky.}
\end{figure*}

But then, something happened---first, I became a dean, and then four
years later I left Carnegie Mellon, as you know. I had a second
administrative career going on the side---actually, I had three
careers, or four. There was also the committee work at the National
Academy, which was a full-time job for awhile, there was the
methodology I worked on in part with students in the Department of
Statistics, and I was also an administrator---I was Department Head for
three years and then I was Dean of the College of Humanities and Social
Sciences. I was on an administrative track in the late 1980s and early
1990s, and my contact with graduate students actually tailed off toward
the end of my time as Dean. I was also teaching, but there are only so
many hours in the day and days in the week. In 1991, I~left and went to
the York University in Toronto as Academic Vice-President (that's like
a provost---they don't have that title at York) and so my regular ties
with graduate students were severed. I resigned from Carnegie Mellon to
go to York, although we didn't sell our Pittsburgh house, and I
returned to Carnegie Mellon a few years later and re-joined the
department.

I like to describe the move back to Carnegie Mellon as a promotion to
the best
position in the univer\-sity---as a tenured professor with no
administrative obligations. I
slowly began to work with graduate students again. Somewhere along
the way I think I~had learned something, which is you can't necessarily get
graduate students to do what you want, and thus what you have to do is
get them to do what they want to do in the best possible way. You
have to get them to complete a thesis, but you have to be able to get
them through and have them gain confidence in what they're doing so
that they think they can make a difference. And I was lucky---I just
had fabulous students; they were terrific people and all the rest of
the stuff just sort of happened. I~had the opportunity to give away
in marriage one of my students, Stella Salvatierra, who was working in
Spain, at a ceremony in the mayor's office in Bilbao, because her
father had a heart attack and couldn't come to the wedding. And there have
been several other weddings since! Because my
students have been so great, the best thing I can do in some
sense is to get them to do the things that they do best. That's in
many ways a serious part of my legacy.\looseness=1

\noinbf{JT:} I was going to ask you what advice you would have
for graduate students in statistics, or undergraduates for that
matter. Clearly, the best advice I could give would be for them to come
to be your students, but since you can't spread yourself totally thin,
failing that, what alternative advice would you offer?

\noinbf{SF:} Well, I really can't work with them all! It's
really bad because now we've got this undergraduate program with
upwards of 150 majors. I can deal with one or two graduate students at
a time. But my advice to budding statisticians is simple: statistics
is an exciting field. There are all these neat problems. There are
neat theories, neat methods, neat applications; we're in a new world.
Big, big data sets. My joint appointments are now in the Machine
Learning Department and in the Heinz College (of Public Policy and
Management). I'm working with data sets that people couldn't conceive
of dealing with a few years ago. And the students I'm working with\vadjust{\goodbreak}
have the ability to go and do things with those data sets that were
unimaginable a decade ago. So my advice is simple. Work with data,
take problems seriously, but you have to learn the mathematics and
statistical theory if you want to do things right. And then you need
to take seriously teaching people what you've done, not just doing the
research. You need to get the descriptions of your work into a form
that other people can understand---that's a really important part of
what we do. That's what National Academy reports are all about. Academy
reports don't have impact if they're badly written. Enormous effort
goes into the executive summaries of reports, into the review process,
and everything up the line. Learning how to do that as a student is
time well spent. It's too late when you're a full professor and you
still haven't learned how to write articles so that other people can
understand what you've done.

%f16 #&#
%
\begin{figure}

\includegraphics{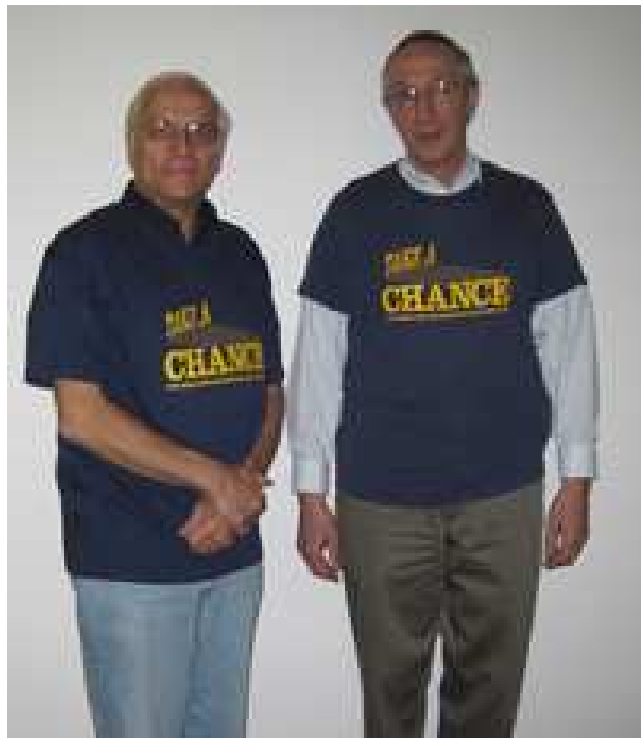}

\caption{Steve and Bill Eddy celebrating the 20th anniversary of \textit
{Chance}, a magazine they co-founded in 1988, wearing their original
Chance t-shirts.}
\end{figure}

%f17 #&#
%
\begin{figure}

\includegraphics{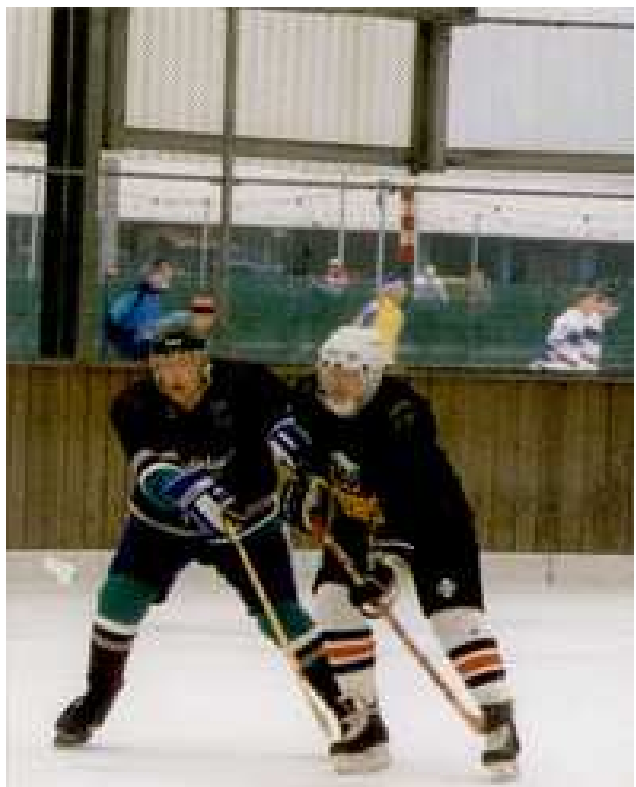}

\caption{Steve (on the right) playing for the Division C national
championship as a member of the Leiden Beaver Beer Team, in Eindhoven,
March, 1997.}
\end{figure}

\noinbf{MS:} So, of your vast experiences, what are you the
most proud of?

\noinbf{SF:} I'm actually proud of a number of things. By the
way, I didn't tell you what my fourth career was. I play ice\vadjust{\goodbreak}
hockey---I still play, that's number one, although the one for which I
have the fewest skills or accomplishments.

\noinbf{MS:} All right, let me interrupt you\ldots

\noinbf{SF:} Ha ha, no-no, as I left the locker room last
Saturday night, one of the guys across the dressing room said to me,
``So how many years have you been playing?'' And I said, ``62.'' He then
said, ``\textit{62}?'' and silence ensued. But maybe hockey is really
number two; number one is
my children and my grandchildren. They're really amazing. They're
another part of my life. Joyce and I were really fortunate; I have two very
smart sons, Anthony and Howard. They have independent careers, they
have lovely wives\ldots

\noinbf{MS:} Where are they now?

%f18 #&#
%
\begin{figure}

\includegraphics{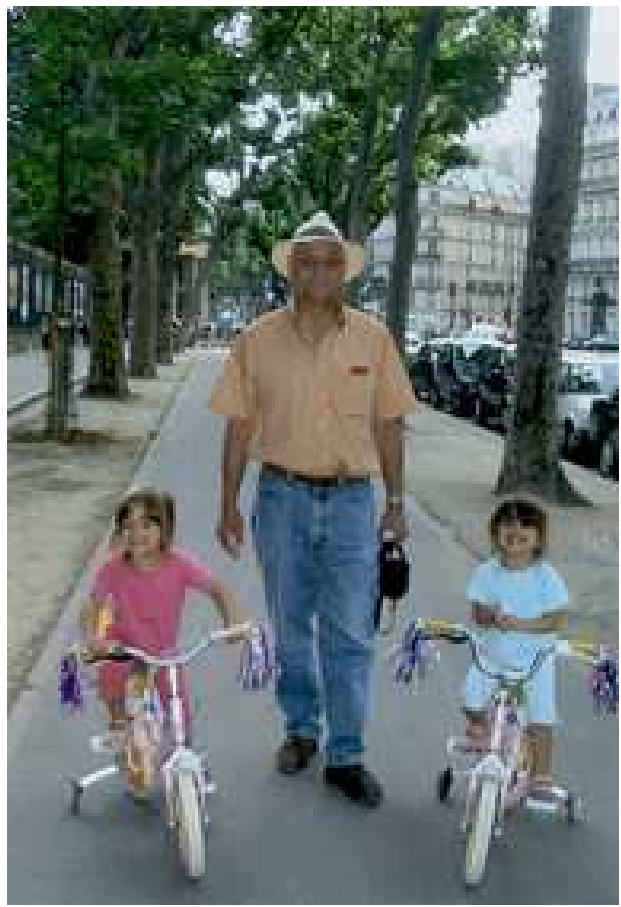}

\caption{Steve with twin granddaughters, Tiffany and Selena, trying out
their new bikes, Paris, 2006.}
\end{figure}

\noinbf{SF:} Anthony lives in Paris, and I have five
grandchildren in Paris, four granddaughters and a grandson. And
Howard lives in the DC area and I have a lovely granddaughter in
Vienna, Virginia. Howard actually has come very close to statistics,
as government liaison for a consortium dealing with surveys and
marketing. The grandchildren are terrific. I love being with them.
We get to look after them every once in a while.

Then there are my students. They're really the people who
are going to do the things that I can only imagine. As I look back
over what I've done, I~see a changed field of statistics. Fred
Mosteller and Bill Kruskal were fabulous---and we've talked about how
they shaped all three of \textit{our} careers, not just my career. And
they launched the Statistics Departments at their respective
universities. I was part of both departments and their programs in
retrospect look ``traditional.'' They emphasized mathematical
statistics and probability. I like to think that when I left
Chicago and went to Minnesota, I started to change what statistics did
and how we thought about it. And applications today sit at the core
of much of statistical theory and methods, and in my department at
Carnegie Mellon our students come out having worked on multiple
applied projects, and they're in demand, because that's the future of
our field. People recognize that advances in statistical
methods---and theory---are intertwined with real\break problems, major
applications. I like to think that I~contributed to the change that
we've seen over the past 40 years.

\noinbf{MS:} Very nice, Steve. What you talk about is a
legacy, not the individual research that may wane in importance over
the years\ldots

\noinbf{SF:} And it's not just my work, it's a collective\ldots\vadjust{\goodbreak}

\noinbf{MS:} But it's the influence of your students, as well
as your children. I wanted to interrupt, because I never thought you
had four careers, I thought you had dozens of careers. You talked
about these professors that, you know, worked $24/7$, so that was your
model. As long as I've known you, you're always multi-tasking, and
you were doing that before the word was even in vogue. You're
fielding questions at a seminar or flying a hockey puck across the
ice. Did any of that rub off on your sons, on your students?

\noinbf{SF:} I don't think that either Anthony or Howard is
quite as obsessed as I am with doing so many things simultaneously.

\noinbf{MS:} How fortunate\ldots

%f19 #&#
%
\begin{figure*}

\includegraphics{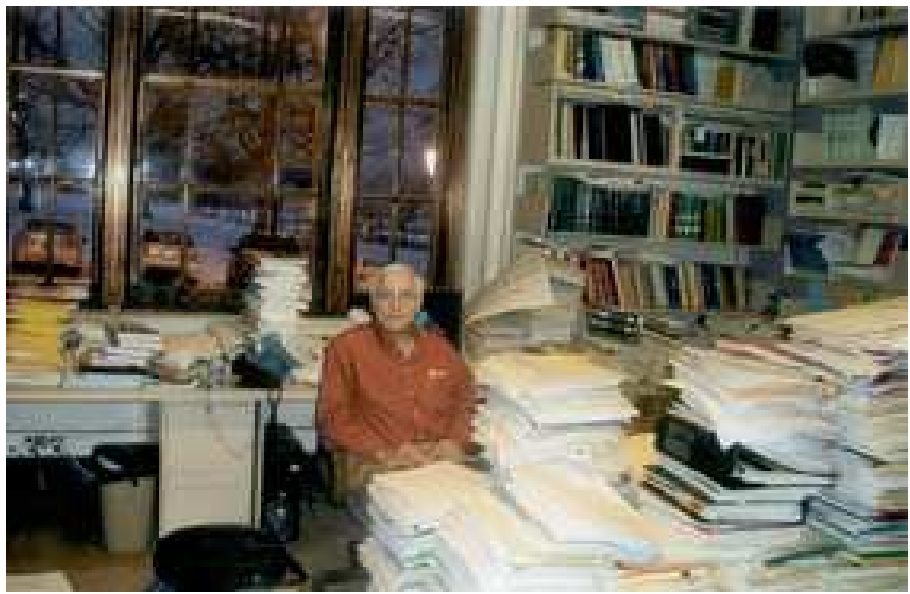}

\caption{Steve, buried amidst files, in his CMU office, 2005.}
\end{figure*}

\noinbf{SF:} That's right! But Anthony did play hockey in Paris for
many years,
and both Anthony and Howard have these terrific
kids---since Anthony has five they take up more of his time than mine
did. Actually, Anthony has inherited some of this multi-tasking, at
least at some
level. He's created his own business in France---a subsidiary of a
Dutch insurance company. His job went from finding
the location to organizing the offices, to hiring the staff, to
inventing the
insurance policies and making sure that they were consistent with the
ones of the parent company.

My students also develop multiple facets of their careers and lives.
I tell them when they come in and
ask if they can work with me that there are a couple of things that
are going to happen if the arrangement is going to succeed. One is
they're going to live and breathe statistics. I see it everywhere.
One of my favorite examples in my little contingency table book came
out of the program from the symphony at the Minneapolis Orchestra one
night when we were there in the 1970s. It didn't \textit{quite} look like
a contingency table, but I made it into one, as Table 2.4. Then in my
book, I described why you shouldn't analyze it the way you would have
otherwise because the units of observation are not independent. At any
rate, I tell the students that I expect them to live and breathe
statistics. They'll get their ideas in the shower\ldots they'll play hard
too, but when all is said and done, if they're not into what they're
doing, they should find another advisor, because other people have
different attitudes about work and how to get your inspiration!
Students of course have their own lives, and as I've said, you don't
tell students what to do, they tell you what they want to do.

\noinbf{JT:} What's next? For you?

\noinbf{SF:} Wow. I'm too busy to stop at the moment to find
out! I still have more than one job. I'm editing, with some others,
the \textit{Annals of Applied Statistics}, I have launched the \textit
{Journal of
Privacy and Confidentiality}, I'm co-chair of the Report Review
Committee at the Academy.\footnote{Steve took over as editor-in-chief
of the
\textit{Annals of Applied Statistics} on January 1, 2013, and is
simultaneously serving
as the founding editor of yet another publication, \textit{The Annual
Review of Statistics and its
Application}, scheduled to launch within the year.} I have a whole
bunch of new Ph.D.
students and post-docs. We've got some absolutely fantastic projects
going on:
research on confidentiality problems and on network modeling, which by
the way, links to confidentiality. Judy and I~also have a book
on surveys and experiments to polish up for publication, as Fred
Mosteller would say.
I have six chapters that were, I had thought, pretty polished at one
stage, but
they are still in a drawer in my office. At least I know where the
drawer is.

\noinbf{JT:} And I know where my copies are\ldots

\noinbf{SF:} And so, I've got more books to write too---with
good collaborators.

\noinbf{MS:} Well, we're almost out of time, but I have one
final question. How would you like to be remembered, Steve?

\noinbf{SF:} Unfortunately not as a great hockey player. As long as my teammates
just let me on the ice, I'm happy to be able to skate around and
get off safely.

I guess I'd like to be remembered as somebody who
produced really good students and who helped change the image of
statistics in the sense that lots of people now work on serious
applied problems and help solve them. And that's not just about
statistics, that's real interdisciplinary scientific work, and that's
the legacy I~inherited from Fred and Bill Krus\-kal and Paul Meier, and
all those other great people that I had a chance to work with, like
Bill Cochran. I would just like for people to think of me in their
kind of company, in some way or another. I suspect that a couple
of decades from now, if anybody ever looks at the video we're making
or reads this interview, they may not remember log-linear models for
contingency tables and other forms of counted data because there will
be new methodology, like the mixed membership and related models
I~now work with. What I know from students today is that, if it wasn't
in the journals in the last three years, they're not sure it's worth
their attention. So, if I~am to have a legacy it needs to be something
larger. I have no
theorems, well, I do have theorems, but none of them are named \textit
{Fienberg's Theorem}. And even if there were a Fienberg's Theorem,
it probably wouldn't be important---what's
important is the attitude, for what statistics is and how it's
recognized by other people outside of our field.

\noinbf{MS:} Well, you've changed statistics, and you've made it fun
along the way. Thank you very much.

%suskaldyti doi

\end{document}